**Engineered Josephson Parametric Amplifier in quantum two-modes squeezed radar**

Seyed Mohammad Hosseiny[1], Milad Norouzi[2], Jamileh Seyed-Yazdi[2]†, Mohammad Hossein Ghamat[3]

[1] Physics Department, Faculty of Sciences, Urmia University, Urmia, Iran

[2] Physics Department, Faculty of Sciences, Vali-e-Asr University, Rafsanjan, Iran

[3] School of Electrical and Computer Engineering, Shiraz University, Shiraz, Iran

† Corresponding author. Email: J.Seyedyazdi@gmail.com

**Abstract**

Josephson parametric amplifier (JPA) engineering is a significant component in the quantum two-mode squeezed radar (QTMS), to enhance, for instance, radar performance and the detection range or bandwidth. In this study, we apply quantum theory to a research domain focusing the design of QTMS radar. We apply engineered JPA (EJPA) to enhance the performance of a quantum radar (QR). We investigate the correlation between the signal and idler using and we propose strategies for maintaining entanglement at room temperature. We define the quantum signal-to-noise ratio (SNR) and detection range equations of the QTMS radar. The engineering JPA, leads to a remarkable improvement of the quantum radar performance, i.e. a large enhancement in quantum SNR of about 6 dB, a substantial improvement in the probability of detection through far fewer channels, and a huge increase in QTMS radar range, from half a meter in the conventional JPA to 482 m in the current study.

**Keywords**: JPA, Engineered JPA, Quantum radar, Quantum SNR, Quantum correlation.

# I. INTRODUCTION

In general, radars transmit radio waves to one or more targets using transmitter antennas and receive and measure echoes using receiver antennas to detect the presence or absence of targets by a detector. Many factors, such as noise and clutter, can be mentioned that call into question this simplicity [1]. The difference between a quantum radar (QR) and a classical radar (CR) can be deduced even from their names. Basic quantum concepts such as the principle of uncertainty, entanglement, correlation, photon statistics, vacuum fluctuations, and squeezing are present in QRs [1-19]. The discussion about QRs has flourished for several years. Several research teams, including Balaji et al. [1-3], Barzanjeh et al. [15], etc., have implemented operational prototypes of QR. All the results obtained by these teams show a significant improvement in the performance of QR compared to the CR counterpart [2-9, 15, 19]. A 6 dB improvement in transmission power, 4 to 6 dB enhancement in SNR, and an improvement of 6 dB in the QR false-alarm compared to CR can be considered as established results in these articles [2-9, 15, 19, 20]. The quantum illumination (QI) range, that gives the accuracy of the square of the mean of the delay of range, can be tens of dB above that of a CR counterpart with the same bandwidth and transmitted energy [20]. In the QR, the samples are much fewer than in the CR, and the signal in QR has a higher correlation coefficient than its classical counterpart [2-9, 15, 19]. The QRs can also be made impervious to hackers by using quantum cryptography [9]. This means that it is possible to create a secure channel by encoding the transmitted photons to protect the information against eavesdropping [9]. The target may also be more visible by using QR rather than CR, due to a quantum effect on the radar cross-section [21]. One type of QR is the quantum two-mode squeezed (QTMS) radar, which is very similar to conventional noise radars [1-3]. These radars use the Josephson parametric amplifier (JPA) and can produce the signal and idler directly in the microwave [1-3, 6, 8, 11, 12, 15, 16]. On the other hand, one of the current disadvantages of QTMS radars is their very high costs of implementation and equipment [1-3, 6, 8, 11, 12, 15, 16].

In recent articles [1-3, 6, 8, 11, 12, 15] it was observed that JPAs have limitations such as low bandwidth and therefore, we need to engineer them to improve the performance of the QR. Engineering JPAs in QTMS radars, hence, gives us the capabilities to build high-range QTMS radars. One of the most important issues for engineers is the range measurement of a QR due to the many inconsistencies in the problem. Therefore, in this study, the range equation of a QTMS radar is introduced and the results are reviewed. We design a QTMS radar with larger bandwidth, better detection range, and improved SNR, taking into account the prototype of the QTMS radar implemented in [1, 3, 15] and using the EJPA [22].

In the current study, after introducing QR and the basic principles of the QTMS radars, we present and evaluate the EJPA and investigate the radar's design using it. Finally, after post-processing the results are presented to show and confirm the capability of our design.

## II. PRELIMINARIES

### A. QUANTUM RADAR (QR)

The basis of the work of a QR can be summarized as follows [1-19]:

1. Using a pump and a signal generator, we produce a current of entangled photon pairs (signal/idler) by quantum sources.

2. To send a signal to the target, we need to amplify the signal with low-noise amplifiers, and to determine the presence of the target, we have to record the idler.

3. After receiving the signal reflected from the target by the receiver antenna, the signal and idler are amplified again, and apply a match-filtering between the received signal and the previously recorded idler.

4. Using a suitable detector, the presence or absence of a target can be inferred. Fig. 1 depicts the general block diagram of a quantum radar.

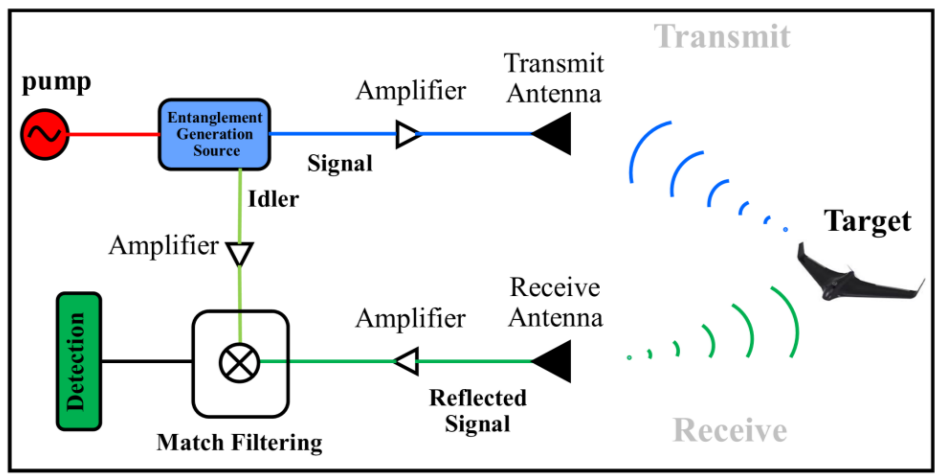

Fig. 1. Schematic block diagram of a QR.

1) **QTMS RADAR**

The template is used to as mentioned, a type of QRs is the quantum two-mode squeezed (QTMS) radar, the operational prototype of which was introduced in [1-3, 5, 6, 8, 15]. Here, the term squeezed refers to the electromagnetic field state that decreases the uncertainty of one component of the field relative to the coherent state (uncertainty in the amplitude and phase of the electric field are the same), increasing uncertainty in the other component [1, 3]. The wave packet, in another words, is compressed or squeezed inside a potential well [23], or, to be more precise, quantum noise decreases in linear compounds of some of the quadrature and increases in other compounds [1-3, 5, 6, 8].

The main part of QTMS radars is the source of entanglement generation, the Josephson parametric amplifier (JPA). JPAs are devices that generate two-mode squeezed vacuum (TMSV) state [1-3, 5, 6, 8]. The term vacuum here refers to vacuum or zero-point energy-related fluctuations, which we intend to amplify. These fluctuations are not similar to classical ones [1, 3]. When the field is quantum, in an absolute vacuum, particles can fluctuate, and these fluctuations lead to spontaneous emission from an excited state at energy level $E_2$ to the ground state at energy level $E_1$, thus emitting a photon with an energy equal to the difference between these two levels, $E_2 - E_1 = \hbar \nu$ [1-3, 5, 6, 8, 23]. JPAs are placed in dilution refrigerators for two reasons: first, because they have a resonant cavity with a superconducting quantum interferometer (SQUID) and superconducting properties, and second, to prevent noise absorption in the entangled signal [1-3, 5, 6, 8, 15]. Fig. 2 shows a schematic representation of the JPA.

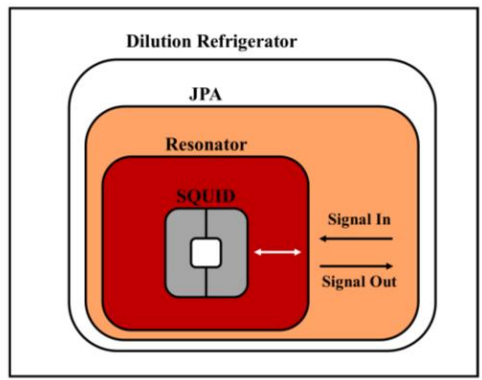

Fig. 2. Schematic representation of a JPA in a dilution refrigerator.

### B. TWO PHOTONS ENTANGLEMENT

A covariance matrix is a matrix whose elements show a correlation between different system parameters. A special type of correlation is entanglement [24-26]. When two beams of light are entangled, they have very strong correlations. The correlations power resulting from quantum entanglement simply cannot occur in classical physics [1, 3]. The root of entanglement is in the quantum superposition principle and has no classical counterpart [26]. In general, in a quantum state, if the measurement on the first qubit affects the result of the measurement of the second qubit, we have an entangled state. Otherwise, it is non- entangled [24, 26].

Since the photons of the signal and idler originate from the same pump photon, there is a strong quantum correlation between the signal and idler, resulting in the squeezing of conjugate signals I and Q [1-3, 5, 6, 8, 15]. For better detection and measurement with classical instruments, the signal and idler need to be amplified in stages. Unfortunately, amplification adds a lot of noise and weakens the entanglement. Entanglement can easily be eliminated by factors such as loss (e. g. antenna gain) and noise (e. g. the presence of amplifiers) [13, 27-29]. The difference between QRs is in the source of their Entanglement. There are different types of entanglement, according to which the source of entanglement production of QRs is also diverse. For example, we have polarization entanglement [7, 19], number of photons [19], superconductivity (Josephson junctions) [1-3, 5, 6, 8, 15], squeezed light [1-3, 5, 6, 8], quantum dot [29 and 28] and etc. There are also two types of continuous-variable entanglement such as voltage and discrete variable such as polarization [1-3, 5, 6, 8, 15]. In QTMS radars, the entanglement of continuous variables of light squeezed by the JPA entanglement generation source is used [1-3].

### C. THE COVARIANCE MATRIX FOR QTMS RADARS

The Hamiltonian of two-modes photons is equal to [23]:

$$\mathcal{H} = i\hbar \left( g a_1^\dagger a_2^\dagger - g^\star a_1 a_2 \right) \tag{1}$$

where subscripts 1 and 2 refer to the two modes. $g$ is the coupling coefficient, $a$ and $a^\dagger$ are the photon annihilation and creation operators, respectively. The two-mode squeezing operator is also expressed as follows [23, 30]:

$$S(\xi) = exp\left(-\xi^* a_1 a_2 + \xi a_1^\dagger a_2^\dagger\right) \tag{2}$$

where $\xi = r\,exp(i\phi)$ is an arbitrary complex number. $r$ and $\phi$ are the amplitude (squeezing parameter) and the phase (squeezing angle), respectively. Also, we have [23, 30]:

$$S^\dagger(\xi) = exp\left(\xi^* a_1 a_2 - \xi a_1^\dagger a_2^\dagger\right) = S^{-1}(\xi) \tag{3}$$

Using the Baker–Campbell–Hausdorff (BCH) formula:

$$e^A B e^{-A} = B + [A,B] + \frac{1}{2!}[A,[A,B]] + \ldots \tag{4}$$

where A and B are two arbitrary operators, and having [30]:

$$S^{-1}(\xi)\left(a_1^\dagger, a_1, a_2^\dagger, a_2\right)^T S(\xi) = M_\xi \left(a_1^\dagger, a_1, a_2^\dagger, a_2\right)^T \tag{5}$$

Thus we have:

$$S^{-1}(\xi) a_1^\dagger S^\dagger(\xi) = a_1^\dagger coshr + a_2 e^{-i\phi} sinhr \tag{6}$$

$$S^{-1}(\xi) a_1 S^\dagger(\xi) = a_1 coshr + a_2^\dagger e^{i\phi} sinhr \tag{7}$$

$$S^{-1}(\xi) a_2^\dagger S^\dagger(\xi) = a_1 e^{-i\phi} sinhr + a_2^\dagger coshr \tag{8}$$

$$S^{-1}(\xi) a_2 S^\dagger(\xi) = a_1^\dagger e^{i\phi} sinhr + a_2 coshr \tag{9}$$

According to the matrix representation of the photon annihilation and creation operators, we obtain [30]:

$$M_\xi = \begin{pmatrix} coshr & 0 & 0 & e^{-i\phi}sinhr \\ 0 & coshr & e^{i\phi}sinhr & 0 \\ 0 & e^{-i\phi}sinhr & coshr & 0 \\ e^{i\phi}sinhr & 0 & 0 & coshr \end{pmatrix} \qquad (10)$$

If $r$ is real, the transformation of Eq. (5) into the position-momentum space is:

$$S^{-1}(\xi)(x_1, x_2, p_1, p_2)^T S(\xi) = L(x_1, x_2, p_1, p_2)^T \qquad (11)$$

with

$$L = H.diag(e^{-r}, e^r, e^r, e^{-r}).H^T \qquad (12)$$

and

$$H = \frac{1}{\sqrt{2}} \begin{pmatrix} 1 & 0 & -1 & 0 \\ 1 & 0 & 1 & 0 \\ 0 & 1 & 0 & -1 \\ 0 & 1 & 0 & 1 \end{pmatrix} \qquad (13)$$

This expresses that $S^\dagger(x_1 - x_2)S = e^{-r}(x_1 - x_2)$ and $S^\dagger(p_1 + p_2)S = e^{-r}(p_1 + p_2)$ means $(x_1 - x_2)$ and $(p_1 + p_2)$ are squeezed. For $\phi = 0$, the covariance matrix will be as follows [30-32]:

$$Cov_{sq2}^{\phi=0} = H^T.diag(e^{-2r}, e^{2r}, e^{2r}, e^{-2r}). \qquad (14)$$

$$H = \begin{pmatrix} cosh2r & 0 & sinh2r & 0 \\ 0 & cosh2r & 0 & -sinh2r \\ sinh2r & 0 & cosh2r & 0 \\ 0 & -sinh2r & 0 & cosh2r \end{pmatrix} \qquad (15)$$

For $\phi \neq 0$, we have [1-3, 5, 6, 8, 31, 32]:

$$Cov_{sq2}^{\phi \neq 0} = \begin{pmatrix} cosh2r & 0 & sinh2rcos\phi & sinh2rsin\phi \\ 0 & cosh2r & sinh2rsin\phi & -sinh2rcos\phi \\ sinh2rcos\phi & sinh2rsin\phi & cosh2r & 0 \\ sinh2rsin\phi & -sinh2rsin\phi & 0 & cosh2r \end{pmatrix} \quad (16)$$

that can also be written as [1-3, 5, 6, 8, 10, 31, 32]:

$$Cov_{sq2}^{\phi \neq 0} = \begin{pmatrix} \sigma_1^2 & 0 & \rho\sigma_1\sigma_2 cos\phi & \rho\sigma_1\sigma_2 sin\phi \\ 0 & \sigma_1^2 & \rho\sigma_1\sigma_2 sin\phi & -\rho\sigma_1\sigma_2 cos\phi \\ \rho\sigma_1\sigma_2 cos\phi & \rho\sigma_1\sigma_2 sin\phi & \sigma_2^2 & 0 \\ \rho\sigma_1\sigma_2 sin\phi & -\rho\sigma_1\sigma_2 sin\phi & 0 & \sigma_2^2 \end{pmatrix} \quad (17)$$

with $\rho\sigma_1\sigma_2 = sinh2r$, $\sigma_1^2 = \sigma_2^2 = cosh2r$ and $\rho = \dfrac{sinh2r}{\sqrt{cosh2r}\sqrt{cosh2r}} = tanh2r$ is the Pearson correlation coefficient.

### III. RESULT AND DISCUSSION

#### A. ENGINEERED JPA (EJPA)

The JPAs are commonly used as a narrowband signal amplifier, meaning that they have limitations (i.e. narrow bandwidth) that prevent the improvement of their performances [1-3, 5, 6, 8, 15, 22].

Our EJPA is similar to that described in [22], where a broadband EJPA by the pumped flux impedance method is presented. Therefore, we use it to construct the quantum radar. One of the advantages of this JPA is the wide bandwidth at low gain rates [22]. By matching the impedance with the input amplifier, its bandwidth is significantly increased from 1 MHz to 300 MHz. The input signal is reflected as an amplified output signal with a gain of about 20 dB [22].

Fig. 3 shows a schematic representation of the equivalent circuit of an EJPA device, in which the entire device is fabricated integrally on intrinsic silicon (>10 kΩcm resistivity). The device operates in a dilution refrigerator with a base temperature of 7 mK [22]. A SQUID loop is made with two Josephson junctions placed in parallel on each side. If the flux line on the chip is combined with two Josephson junctions, flux pumping is provided. The λ/4 resonator with a characteristic impedance of 45 Ω reduces the JPA resonant quality factor. On the other hand, the λ/2 resonator with a characteristic impedance of 80 Ω reduces the frequency dependence in the system sensitivity matrix, and this leads to amplification of the bandwidth [22].

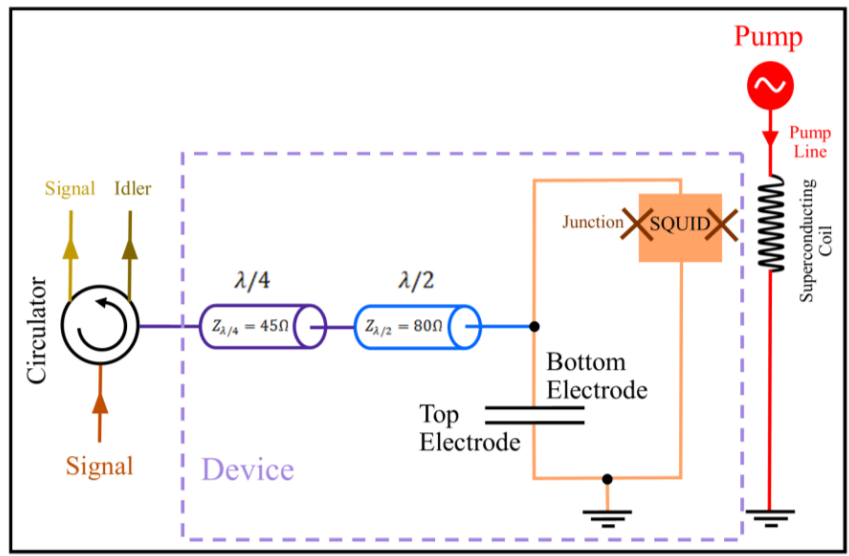

Fig. 3. Schematic representation of an EJPA circuit.

A parallel plate capacitor with two top and bottom electrodes with a total input capacity c = 2.03 ± 0.02 pF is located at the JPA input which is connected to the ground. The input line is galvanically connected to the lower electrode by impedance, which is directly connected to the SQUID. The upper electrode of the capacitor is connected to the ground in parallel with the SQUID with a non-galvanic connection. By connecting galvanic to the input, the coupling quality coefficient decreases, and therefore, the amplification bandwidth increases [22].

The JPA used in the current study, is the degenerate four-wave type which means that, the input and output frequencies are identical. The signal (and idler) frequency here is 5.31 GHz [22].

### B. QTMS RADAR DESIGN WITH EJPA

The block diagram of QTMS Radar is illustrated in Fig. 4.

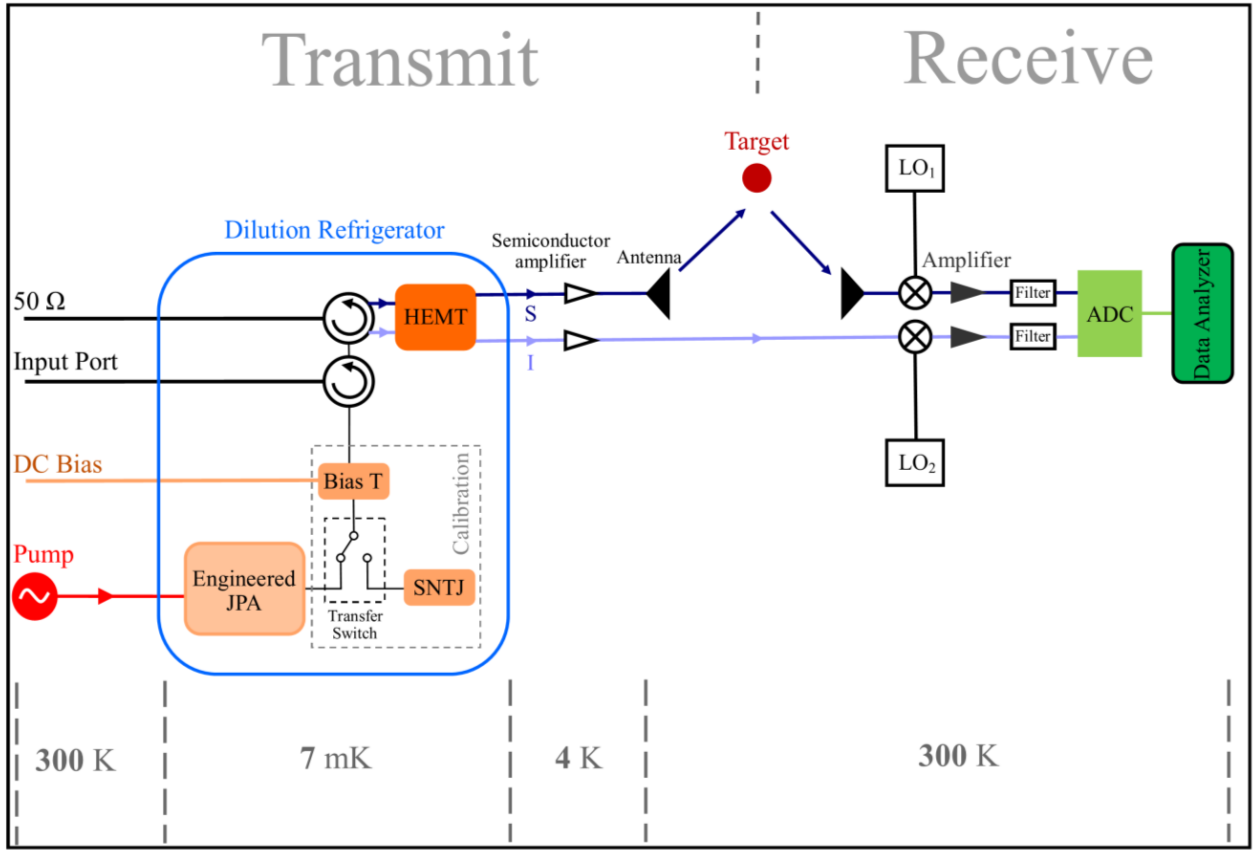

Fig. 4. Block diagram of the QR using EJPA.

The JPA bandwidth is increased from 1 MHz [1, 3] by the pumped flux impedance method, to 300 MHz. The pump power is 5 dBm. The JPA is located inside the refrigerator, as shown in Fig. 4, is connected to the T-bias via a microwave switch (which can turn the pump on or off) and a device called a Shot-Noise Tunnel Junction (SNTJ). These are used as part of the calibration process to confirm the entanglement of the JPA output signal [1-3, 5, 6, 8]. The output signal is amplified to make it easier to measure and detect. Because amplifiers add noise to the signal, and usually the added noise is from the first amplifier, we use high-electron-mobility transistor (HEMT) (which is a low-noise amplifier) and a semiconductor amplifier [1-3, 5, 6, 8, 15]. HEMT is also placed in a dilution refrigerator and semiconductor amplifiers operate at 4K. The refrigerator used here is the Bluefors LH250, which uses liquid helium for cooling. The cooling power of the refrigerator for a temperature of 7mK is about 10 μW [31].

After calibration, circulators are placed, which act as insulators for our system to prevent additional signal and noise from reaching the JPA [1-3, 5, 6, 8, 15]. After amplification, signal is sent to the target by the C-band antenna, but we record the idler. It should be noted that no measurements are made on the idler before the signal arrives. The local oscillators (LO) $LO_1$ and $LO_2$ are applied to the reflection signal of the target and the idler to convert the frequency to the intermediate frequency (20 MHz) [15], because the processing of the signal with a fixed frequency enhances the performance of the receiver. Finally, the signal is amplified again and detected after digitization.

### C. POST-PROCESSING

The idler and signal modes are recorded after amplification with a dual-channel ADC AD570JD with 8-bit resolution [15]. We have to do the match-filtering between the received signal and the signal that is recorded inside the system [1-3, 5, 6, 8, 15, 34].

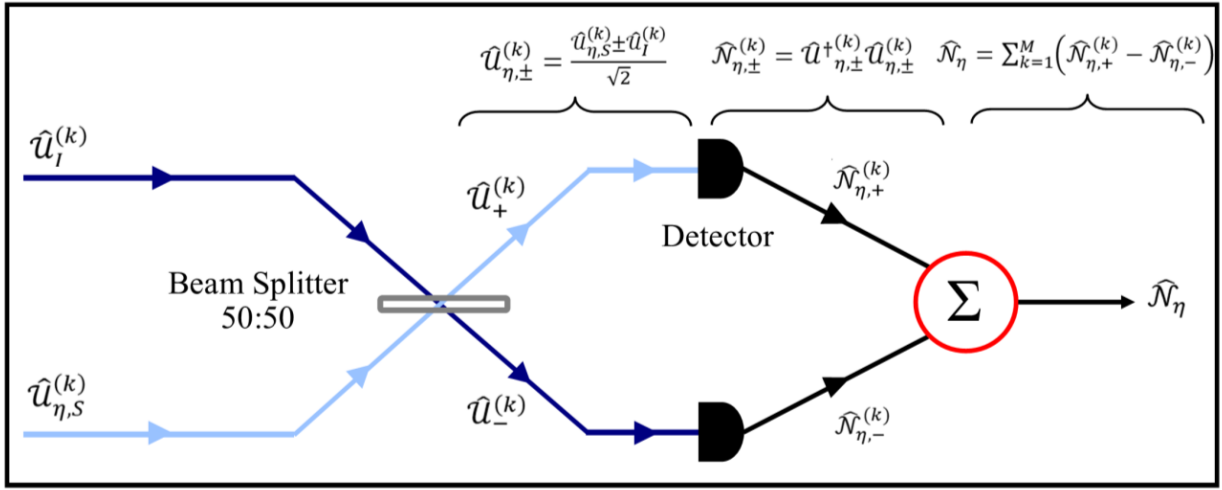

Fig. 5. The EJPA quantum post-processing schema.

According to Fig. 5, $\hat{\mathcal{U}}_{\eta,S}^{(k)}$ and $\hat{\mathcal{U}}_{I}^{(k)}$ are mixed by a 50:50 beam splitter and the output is as follows [15, 34]:

$$\hat{\mathcal{U}}_{\eta,\pm}^{(k)} = \frac{\hat{\mathcal{U}}_{\eta,S}^{(k)} \pm \hat{\mathcal{U}}_{I}^{(k)}}{\sqrt{2}} \qquad (18)$$

After detecting these modes, the photon counts are equal to the measurements of the corresponding number operator [15, 34]:

$$\hat{\mathcal{N}}_{\eta,\pm}^{(k)} = \hat{\mathcal{U}}_{\eta,\pm}^{\dagger(k)} \hat{\mathcal{U}}_{\eta,\pm}^{(k)} \qquad (19)$$

The total photon counts for the two detectors are equal to

$$\hat{\mathcal{N}}_{\eta} = \sum_{k=1}^{M} \left( \hat{\mathcal{N}}_{\eta,+}^{(k)} - \hat{\mathcal{N}}_{\eta,-}^{(k)} \right) \qquad (20)$$

According to the recent literature [34, 15], the microwave mode of the signal return to post-processing ($\eta = 0$ in the absence of the target and $\eta \neq 0$ in the presence of the target) is equal to [34, 15]:

$$\hat{\mathcal{U}}_{\eta,S}^{(k)} = \sqrt{G_s}\left[\sqrt{\eta}\hat{a}_S + \sqrt{\frac{\eta(G_s^A-1)}{G_s^A}}\hat{a}_{n,S}^{\dagger A} + \sqrt{\frac{(1-\eta)}{G_s^A}}\hat{a}_n^E + \sqrt{\frac{(G_s^A-1)}{G_s}}\hat{a}_{n,S}^{\dagger D}\right] + \sqrt{2}\hat{a}_V \tag{21}$$

As well, the idler's microwave mode towards the post-processor

$$\hat{\mathcal{U}}_I^{(k)} = \sqrt{G_I}\left[\hat{a}_I + \sqrt{\frac{(G_I^A-1)}{G_I^A}}\hat{a}_{n,I}^{\dagger A} + \sqrt{\frac{(G_I^A-1)}{G_I}}\hat{a}_{n,I}^{\dagger D}\right] \tag{22}$$

Here $G_s = G_s^D(dB)G_s^A(dB)$ is the system signal gain where $G_s^D$ and $G_s^A$ are the detection signal gain and amplification signal gain, respectively. Also $G_I = G_I^D(dB)G_I^A(dB)$ is the system idler gain. The effective dissipation range is $-25\,dB < \eta < 0\,dB$. $\hat{a}_S$ and $\hat{a}_I$ are the signal and idler annihilation operators, respectively. $\hat{a}_{n,S}^{\dagger A}$ and $\hat{a}_{n,I}^{\dagger A}$ are the signal and idler amplification noise creation operators at 4K, respectively. $\hat{a}_n^E$ is the environment noise mode operator at room temperature and $\hat{a}_V$ is the vacuum mode. $\hat{a}_{n,S}^{\dagger D}$ and $\hat{a}_{n,I}^{\dagger D}$ are the signal and idler detection creation operators at 290 K, respectively [15, 34].

The error probability is obtained in two parts corresponding to the absence or presence of the target (the false-alarm probability $P_f$ and the miss probability $P_r$). The error probability in quantum receivers is [15, 34-45]:

$$P_{D(M)}^{QR} = \frac{P_f + P_m}{2} = \frac{erfc\left[\sqrt{(SNR)_M^{QR}/8}\right]}{2} \tag{23}$$

The detection probability in terms of the false-alarm probability for QTMS radar according to previous publications [1-3, 5, 6, 8, 32, 33], is expressed by:

$$P_D^{QR} = Q\left(\frac{\rho\sqrt{2N}}{1-\rho^2}, \frac{\sqrt{-2\ln P_{FA}}}{1-\rho^2}\right) \qquad (24)$$

where $N$ is the number of channels, $Q$ is the Marcom function [34] and $\rho$ is quantum correlation coefficient [35-45]:

$$\rho = \frac{\rho_0}{\sqrt{1+\left(1/SNR^{QR}\right)^4}} \qquad (25)$$

where the $SNR^{QR}$ is the quantum signal to noise ratio for a QTMS radar and is defined as [15, 34]:

$$(SNR)_M^{QR} = \frac{4M\left[\left(\left|\hat{\mathcal{N}}_{\eta,+}\right\rangle_{H_1} - \left|\hat{\mathcal{N}}_{\eta,-}\right\rangle_{H_1}\right) - \left(\left|\hat{\mathcal{N}}_{\eta,+}\right\rangle_{H_0} - \left|\hat{\mathcal{N}}_{\eta,-}\right\rangle_{H_0}\right)\right]^2}{\left(\sqrt{\left\langle\Delta\hat{\mathcal{N}}_\eta^2\right\rangle_{H_0}} + \sqrt{\left\langle\Delta\hat{\mathcal{N}}_\eta^2\right\rangle_{H_1}}\right)^2} \qquad (26)$$

Here $M = B.\tau$ is the number of modes, $B$ is the bandwidth, and $\tau$ is the time. Also, the value expected indicates the average over the total $M$ copy. Therefore, we have [15, 34, 36, 37]:

$$\left\langle\Delta\hat{\mathcal{N}}_\eta^2\right\rangle_{H_1} = \left\langle\hat{\mathcal{N}}_{\eta,+}\right\rangle_{H_1}\left(\left\langle\hat{\mathcal{N}}_{\eta,+}\right\rangle_{H_1}+1\right)+\left\langle\hat{\mathcal{N}}_{\eta,-}\right\rangle_{H_1}\left(\left\langle\hat{\mathcal{N}}_{\eta,-}\right\rangle_{H_1}+1\right) - \frac{\left(\left\langle\hat{\mathcal{U}}_{\eta,S}^{\dagger(k)}\hat{\mathcal{U}}_{\eta,S}^{(k)}\right\rangle_{H_1} - \left\langle\hat{a}_I^\dagger\hat{a}_I\right\rangle\right)^2}{2} \qquad (27)$$

$$\left\langle\Delta\hat{\mathcal{N}}_\eta^2\right\rangle_{H_0} = \left\langle\hat{\mathcal{N}}_{\eta,+}\right\rangle_{H_0}\left(\left\langle\hat{\mathcal{N}}_{\eta,+}\right\rangle_{H_0}+1\right)+\left\langle\hat{\mathcal{N}}_{\eta,-}\right\rangle_{H_0}\left(\left\langle\hat{\mathcal{N}}_{\eta,-}\right\rangle_{H_0}+1\right) - \frac{\left(\left\langle\hat{\mathcal{U}}_{\eta,S}^{\dagger(k)}\hat{\mathcal{U}}_{\eta,S}^{(k)}\right\rangle_{H_0} - \left\langle\hat{a}_I^\dagger\hat{a}_I\right\rangle\right)^2}{2} \qquad (28)$$

$$\left\langle\hat{\mathcal{N}}_{\eta,+}\right\rangle_{H_1} = N_v + \frac{1}{2}\left[G_s\eta(N_s+1) + \frac{G_s\eta\bar{n}_{A,s}\left(G_s^A-1\right)}{G_s^A} + \frac{G_s\left(\bar{n}_E+1\right)(1-\eta)}{G_s^A} + \left(G_s^D-1\right)\left(\bar{n}_{D,s}\right)\right] \qquad (29)$$

$$+\frac{1}{2}G_I N_I + \frac{1}{2}G_I\left(\bar{n}_{Add,I}+1\right) + \frac{1}{2}\sqrt{G_s G_I}\eta\hat{a}_s\hat{a}_I$$

$$\left\langle \hat{\mathcal{N}}_{\eta,-} \right\rangle_{H_1} = N_v + \frac{1}{2}\left[ G_s\eta(N_s+1) + \frac{G_s\eta\bar{n}_{A,s}(G_s^A-1)}{G_s^A} + \frac{G_s(\bar{n}_E+1)(1-\eta)}{G_s^A} + (G_s^D-1)(\bar{n}_{D,s}) \right]$$
$$+ \frac{1}{2}G_I N_I + \frac{1}{2}G_I(\bar{n}_{Add,I}+1) - \frac{1}{2}\sqrt{G_s G_I}\eta\, \hat{a}_s \hat{a}_I \qquad (30)$$

$$\left\langle \hat{\mathcal{U}}_{\eta,S}^{\dagger(k)}\hat{\mathcal{U}}_{\eta,S}^{(k)} \right\rangle_{H_1} = 2N_v + \left[ G_s\eta(N_s+1) + \frac{G_s\eta\bar{n}_{A,s}(G_s^A-1)}{G_s^A} + \frac{G_s(\bar{n}_E+1)(1-\eta)}{G_s^A} + (G_s^D-1)(\bar{n}_{D,s}) \right] \qquad (31)$$

$$\left\langle \hat{\mathcal{N}}_{\eta,+} \right\rangle_{H_0} = N_v + \frac{1}{2}\left[ G_s^D(\bar{n}_E+1) + \bar{n}_{D,s}G_s^D\left(1-\frac{1}{G_s^D}\right) + \frac{1}{2}G_I N_I + \frac{1}{2}G_I(\bar{n}_{Add,I}+1) \right] \qquad (32)$$

$$\left\langle \hat{\mathcal{N}}_{\eta,-} \right\rangle_{H_0} = N_v + \frac{1}{2}\left[ G_s^D(\bar{n}_E+1) + \bar{n}_{D,s}G_s^D\left(1-\frac{1}{G_s^D}\right) + \frac{1}{2}G_I N_I + \frac{1}{2}G_I(\bar{n}_{Add,I}+1) \right] \qquad (33)$$

$$\left\langle \hat{\mathcal{U}}_{\eta,S}^{\dagger(k)}\hat{\mathcal{U}}_{\eta,S}^{(k)} \right\rangle_{H_0} = 2N_v + \left[ G_s^D(\bar{n}_E+1) + \bar{n}_{D,s}G_s^D\left(1-\frac{1}{G_s^D}\right) + \frac{1}{2}G_I N_I + \frac{1}{2}(\bar{n}_{Add,I}+1) \right] \qquad (34)$$

The subscripts 1 and 0 in $H_1$ and $H_0$ represent the presence or absence of target, respectively. Also, $N_s$ and $N_I$ are the signal and idler photons numbers, respectively. $\bar{n}_{Add,I}$ and $\bar{n}_{Add,s}$ are the system noise average of both measurement channels. $\bar{n}_E$ is the average number of environmental thermal photons. $\bar{n}_{D,s}$ and $\bar{n}_{D,I}$ are the average number of signal and idler noises received. $\bar{n}_{A,s}$ and $\bar{n}_{A,I}$ are the photons number average of signal and idler amplification, and $N_v$ is the number of vacuum modes [34, 15].

### D. *SIMULATION*

In this study, we used C-band antennas (4-8 GHz). These antennas have less losses than X-band antennas (8-12 GHz) [1-3]. System parameters are calculated in Table. 1 as follows:

TABLE I

THE CALCULATED PARAMETERS OF QTMS RADAR FOR SIMULATION

| Quantity | EJPA [This work] | JRM [15] | JPA [1] | Unit |
|---|---|---|---|---|
| Antenna | C-band | X-band | X-band | --- |
| Antenna gain (G) | 6.4 | 15 | 15 | dB |
| Antenna effective area ($A_e$) | $8.8 \times 10^{-5}$ | ---- | ---- | $m^2$ |
| Target radar cross section (σ) | 1.0 | 1.0 | 1.0 | $m^2$ |
| Bandwidth (B) | 300 | 20 | 1.0 | MHz |
| JPA power gain ($G_p$) | 20 | 30 | 20 | dB |
| HEMT gain (at 4K) ($G_{HEMT}$) | 38 | 36 | 36 | dB |
| Signal gain ($G_S$) | 83.98 | 93.98 | ~96 | dB |
| Detection gain ($G^D$) | 16.82 | 16.82 | 16.82 | dB |
| Amplifier gain ($G^A$) | 67.16 | 77.16 | 79.18 | dB |
| Pump power ($P_p$) | 5 | -97 | -82 | dBm |
| Noise power ($P_n$) | -145 | 4 | -94 | dBm |
| Pump frequency $\omega_p = \omega_s + \omega_i$ | 10.62 | 16.89 | 13.6821 | GHz |
| **Signal frequency ($\omega_s$)** | 5.31 | 10.09 | 6.1445 | GHz |
| **Idler frequency ($\omega_i$)** | 5.31 | 6.8 | 7.5376 | GHz |
| **Signal-to-noise ratio ($SNR^{QR}$)** | -13.48 | -18 | -19 | dB |
| **Range (R)** <br> ($N_s = 0.1$ and $\eta = 1$ dB ) | 482 | 1.0 | 0.5 | m |

1) *SNR AND ROC (Receiver Operating Characteristic)*

An Using Eqs (26-34), Fig. 6 shows the SNR versus $N_s$ plot comparing the three scenarios: conventional JPA [1,3], Josephson ring modulators (JRM) [15], and EJPA. The clear conclusion that can be deduced from this plot is that the SNR performs better in the EJPA scenario, than in the other scenarios. The SNR for EJPA is about 5 dB better than JRM, and about 6 dB better than conventional JPA. Therefore, the EJPAs are promising to improve SNR, meaning a better performance of quantum radars.

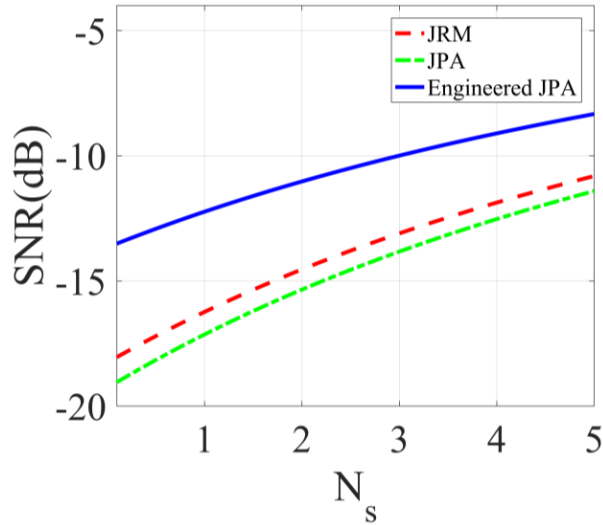

Fig. 6. Comparison of SNRs for conventional JPA, JRM, and EJPA versus $N_s$.

Using Eq. (24), the ROC diagram for different scenarios is depicted in Figs 7 and 8, which clearly show the superiority of the EJPA compared to other scenarios. In Fig. 7, the probability of detection in EJPA is better than other scenarios.

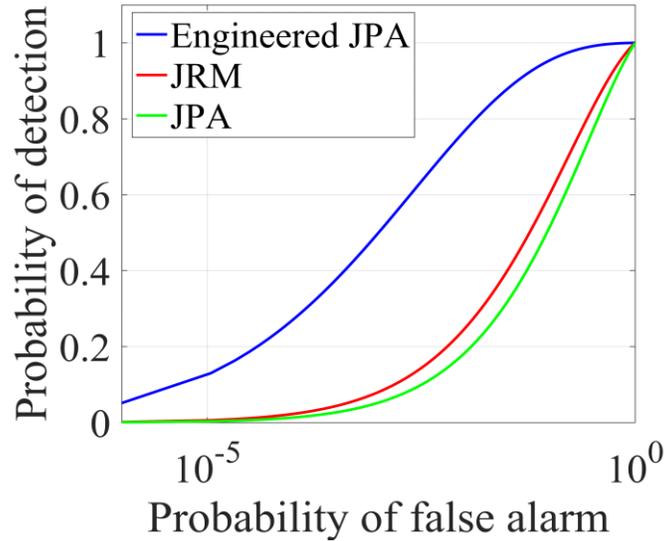

Fig. 7. The ROC comparison plot. Comparison plot for conventional JPA (green), JRM (red) and EJPA (blue), for $N_s$=0.1, $N$=150 and $\rho_0$=1.

The ROC comparison plot for conventional JPA, JRM, and EJPA is illustrated in Fig. 8. It is clear that the detection probability in an EJPA with a smaller number of N channels reaches a maximum of one, demonstrated a significant improvement.

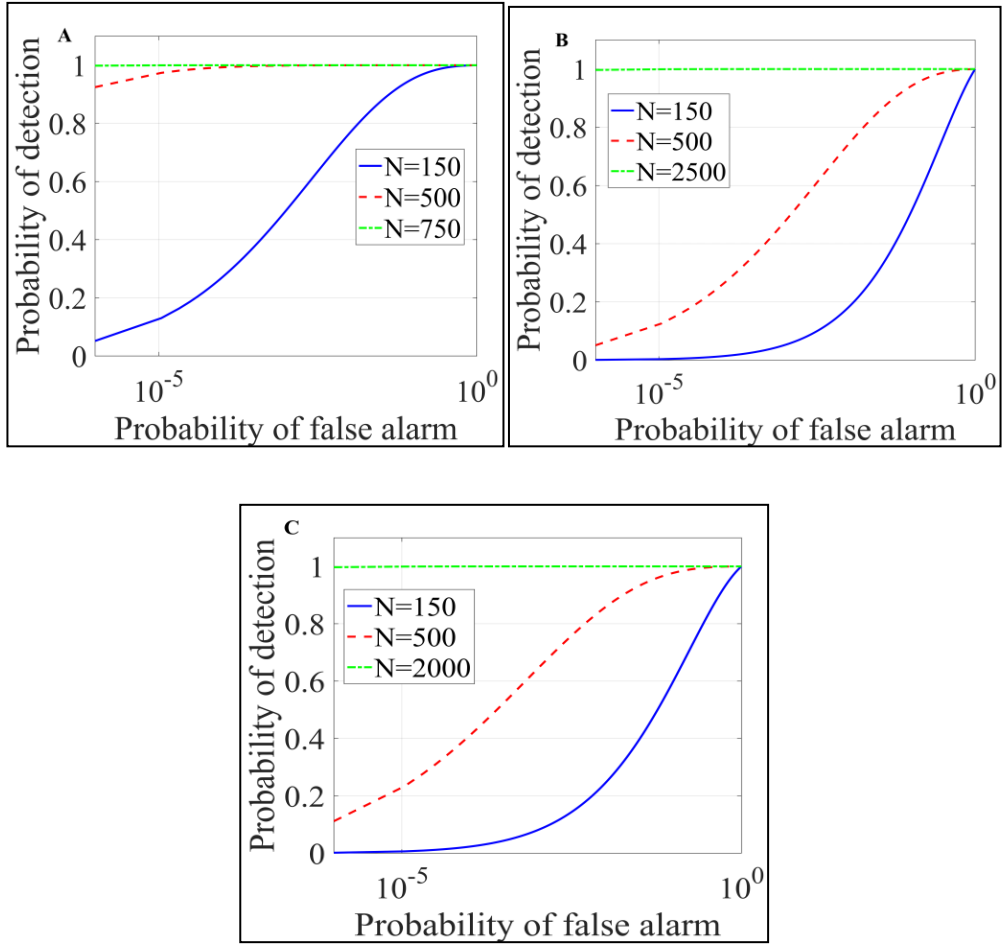

Fig. 8. The ROC comparison plots. (A) conventional JPA, (B) JRM, and (C) EJPA with different channels numbers ($N$).

The SNR plot is also examined versus a correlation function in Fig. 9. This figure shows that when the correlation increases, the quantum SNR also increases. Therefore, the correlation and entanglement play the most important role in QTMS radars and we need to fabricate more correlated signal and idler (by engineering the JPAs) for maintaining entanglement in room [37-45].

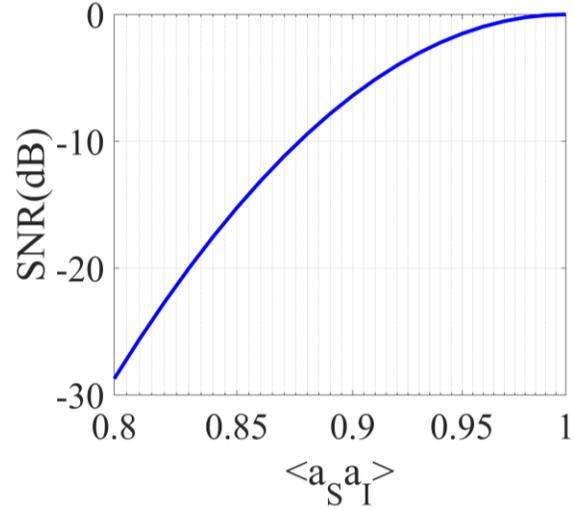

Fig. 9. SNR plot as a correlation function for an EJPA.

### 2) QTMS RADAR RANGE

As mentioned earlier, range evaluation is one of the most important tasks of the quantum radar. Since QTMS radars are very similar to conventional noise radars, the QTMS range equation would be as follows [35, 46, 47]:

$$R_{max} = \left( \frac{GA_e \sigma P_s}{(4\pi)^2 P_n (SNR)^Q_{min}} \right)^{1/4} \quad (35)$$

where $G$ is the antenna gain, $\lambda$ is the wavelength, $A_e = G\lambda^2/4\pi$ is the effective antenna area, σ is the target radar cross-section, $P_s$ and $P_n$ are the signal and noise power respectively. The only difference between this equation and the conventional noise radar equation is $(SNR)^Q_{min}$, which is quantum here. Fig. 10 depicts the detection range versus $(SNR)^Q_{min}$. It is clear that the range in EJPA has a remarkable increase from 0.5 m in conventional JPA [1-3], and 1 meter in JRM [15], to several hundred meters (482 m). Additionally, SNR loses performance efficiency with increasing range.

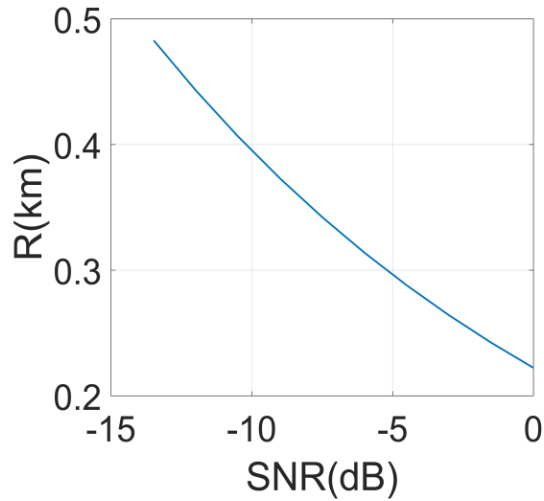

Fig. 10. Detection range versus SNR of EJPA.

## IV. Conclusion

In this study, a QTMS (quantum two-mode squeezed) radar based on an EJPA (engineered Josephson parametric amplifier) quantum source was designed and its performance was evaluated and compared to two other radar scenarios: the conventional JPA quantum source, and Josephson ring modulator (JRM). The application of an EJPA in QTMS radar leads to a significant overall improvement in radar performance.

The correlation between the signal and idler is the most important part of maintaining entanglement at room temperature, and the more the correlation between the two can be increased, the less we will worry about the entanglement suppressing. Therefore, we must fabricate JPAs that can generate signals and idlers with much higher correlations (by engineering the JPAs). From our findings, the quantum SNR shows a performance efficiency of about 5 dB relative to the JRM and about 6 dB relative to the conventional JPA. The detection probability is also remarkably higher than the other two considered scenarios. Moreover, our channel numbers in the detection probability is much less, compared with the other two scenarios. Furthermore, the QTMS radar range equation was defined, and it was expressed that as the QTMS radar range increases, the quantum SNR decreases proportionally. Finally, we obtained a remarkable increase of 482 meters for the QTMS radar range, compared to one meter in the JRM and half a meter in the conventional JPA. Therefore, we confirmed that, using the EJPA in QTMS radar, is very promising to get a considerable improvement in radar performance.

**Acknowledgments**

We wish to acknowledge the financial support of the MSRT of Iran, Urmia University, Vali-e-Asr University of Rafsanjan and Shiraz University.